\documentclass[preprint,showpacs,preprintnumbers,amsmath,amssymb]{revtex4}
\begin{document}

\title{Boundary conditions and amplitude ratios for finite-size corrections
of a one-dimensional quantum spin model}

\author{N. Sh. Izmailian$^{1,2,3,4,*}$ and
Chin-Kun Hu$^{1,5,+}$}

\affiliation{$^1$Institute of Physics, Academia Sinica, Taipei
11529, Taiwan}

\affiliation{$^2$Yerevan Physics Institute, Alikhanian Br. 2,
375036 Yerevan, Armenia}

\affiliation{$^3$ International Center for Advanced Study, Yerevan
State University, 1 Alex Manoogian St., Yerevan, 375025, Armenia}

\affiliation{$^4$National Center of Theoretical Sciences at
Taipei, Physics Division, National Taiwan University, Taipei
10617, Taiwan}

\affiliation{$^5$Center for Nonlinear and Complex Systems and
Department of Physics, Chung-Yuan Christian University, Chungli
320, Taiwan}

\date{\today}

\begin{abstract}
We study the influence of boundary conditions on the finite-size
corrections of a one-dimensional (1D) quantum spin model by exact
and perturbative theoretic calculations. We obtain two new
infinite sets of universal amplitude ratios for the finite-size
correction terms of the 1D quantum spin model of $N$ sites with
free and antiperiodic boundary conditions. The results for the
lowest two orders are in perfect agreement with a perturbative
conformal field theory scenario proposed by Cardy [Nucl. Phys. B
{\bf 270}, 186 (1986)].
\end{abstract}

 \maketitle

\section{Introduction}

%% 99Ising,04prl,05prlwh,BcShape

Universality
\cite{71gene,barber,70fss,ufssf1,ufssf,aharony1,zinn}, finite-size
scaling
\cite{barber,70fss,ufssf1,ufssf,aharony1,zinn,2dperco,3dperco,
96Ising,99Ising,04prl,05prlwh,BcShape}, and finite-size
corrections
\cite{fsc,97ziff,99preFSC,01prl,01FSC,caselle,ih01,ih02,
02jpaFSC,03preDimer,05prlDimer,07sigma,06preDimer} for critical
lattice systems have attracted much attention in recent decades.
%% \cite{fsc,97ziff,99preFSC,01prl,01FSC,caselle,ih01,ih02,ih03,02jpaFSC,
%% 03preDimer,05prlDimer,07sigma,06preDimer,delfino1,delfino2}.
It has been found that critical systems can be classified into
different universality classes so that the systems in the same
class have the same set of critical exponents, whose values
depends only on the global properties of the system such as
spacial dimensions, number of components of the order parameter,
the range of interaction, and the symmetry group
\cite{71gene,barber,70fss,ufssf1,ufssf,aharony1,zinn}. The
hypothesis of universality has much stronger implications and it
is possible to show that models belonging to the same universality
class also share the same set of universal finite-size scaling
functions (UFSSFs) and amplitude ratios
\cite{ufssf1,ufssf,aharony1,zinn,2dperco,3dperco,
96Ising,99Ising,01prl,delfino1,delfino2}, whose values are
independent of the microscopic structure of interactions. However,
using a histogram Monte carlo simulation method \cite{92hu}, Hu,
{\it et al.} confirmed for percolation the well-established
dependence of finite-size scaling functions on boundary conditions
and on the shape of the domains of the systems
\cite{BcShape,2dperco,3dperco,04prl,05prlwh}. Based on exact
partition functions of the Ising model on square, triangular, and
honeycomb lattices \cite{02jpaWu}, Wu, Hu and Izmailian have found
similar results for the Ising model \cite{03preWu}.

%% \bibitem{fsc} A. E. Ferdinand, J. Math. Phys. {\bf 8}, 2332
%% (1967);  A. E. Ferdinand and M. E. Fisher, Phys. Rev.
%% \textbf{185}, 832 (1969).

%% \bibitem{97ziff}  R. M. Ziff, S. R. Finch and V. S. Adamchik, Phys Rev. Lett.
%% {\bf 79}, 3447 (1997).

In 1967-1969 Ferdinand and Fisher \cite{fsc} calculated exact
finite-size corrections (FSCs) for the free energy and its
derivatives of the dimer and Ising models. In 1997, Ziff, Finch
and Adamchik \cite{97ziff} used Monte Carlo simulations to
calculate FSCs for cluster numbers of two-dimensional random
percolation models. In 2002, Caselle, {\it et al.}  \cite{caselle}
used conformal field theory to study correction terms for the free
energy and its derivatives of the Ising model. Based on the
connections between lattice phase transition models and correlated
percolation models \cite{corr-perco}, in 1999 Hu, {\it et al.}
\cite{99preFSC} calculated FSCs for cluster numbers of the
$q$-state Potts model for $q$ being 1, 2, 3 and 4. In such
studies, they extended the calculations of Ferdinand and Fisher
for the Ising model on torus \cite{fsc} to higher orders. Such
studies inspire further research on FSCs for the Ising
\cite{01prl} and the dimer models.

Based on Kronecker's double series and exact asymptotic expansions
for exact partition functions \cite{02jpaFSC}, exact finite-size
corrections for the Ising model \cite{01FSC,ih01,ih02,02jpaFSC}
and the dimer model
\cite{03preDimer,05prlDimer,07sigma,06preDimer} on planar lattices
with various boundary conditions have been obtained.  It has been
found that such correction terms also depend on the boundary
conditions. In \cite{01prl}, we have found exact universal
amplitude ratios for finite-size corrections of the
two-dimensional Ising model on square, plane triangular and
honeycomb lattices and a quantum spin chain on lattices with
periodic boundary conditions. It is of interest to know how
universal amplitude ratios depend on boundary conditions. In the
present paper, we will address this question for a one dimensional
quantum spin model with a varying parameter $\gamma$. We have
calculated exact amplitude ratios for the quantum spin model with
different boundary conditions and found that such ratios depend on
boundary conditions and are independent of $\gamma$.

The quantitative description of the universality classes of
critical behavior is the main goal of quantum field theory when it
is  applied to statistical mechanics. In principle, the universal
quantities can be computed from the quantum field theory encoding
the fundamental symmetries of the system.

The universality class can describe the critical behavior of many
physical systems, which undergo a second order phase transition.
The criteria by which the critical systems can be classified into
different universality classes is a problem of much academic
interest. Two-dimensional (2D) critical systems are parameterized
by the conformal anomaly $c$ which is the central charge in the
Virasoro algebra \cite{Belavin,Dotsenko1,Dotsenko2}. The conformal
anomaly $c$ can be obtained directly from the finite size
corrections to the free energy for a 2D classical system on
infinitely long but finitely wide strip at a conformal  invariant
critical point.

The asymptotic finite-size scaling behavior of the critical free
energy ($f_N$) and the inverse correlation lengths ($\xi_i^{-1}$)
of an infinitely long 2D strip of finite width $N$ at criticality
has the form \cite{Blote,Blote1}
\begin{equation}
\lim_{N \to \infty} {N^2 (f_N-f_{\infty})- 2 N f_{surf}}=A,
\label{I2}
\end{equation}
\begin{equation}
\lim_{N \to \infty} {N \xi_i^{-1}}=D_i, \label{I1}
\end{equation}
where $f_{\infty}$ is the bulk free energy, $f_{surf}$ is the
surface free energy and $A$ and $D_i$ are the universal constants,
but may depend on the boundary conditions (BCs). In some 2D
geometries, the values of $A$ and $D_i$ are known
\cite{Blote,Blote1,cardy86a}, to be related to the conformal
anomaly number ($c$), the conformal weight of the ground state
$(\Delta)$, and  the scaling dimension of the $i$-th scaling field
($x_i$) of the theory
\begin{eqnarray}
A&=& 4\pi \left(\frac{c}{24}-\Delta\right), \qquad D_i = 2 \pi
x_i, \label{Aperiod}
\end{eqnarray}
for periodic or antiperiodic BCs and
\begin{eqnarray}
A&=& \pi \left(\frac{c}{24}-\Delta\right),  \qquad \quad D_i = \pi
x_i, \label{Afree}
\end{eqnarray}
for free BCs. For the 2D Ising model, we have $c=1/2$.  The
conformal weight of the ground state $\Delta$, and  the scaling dimension $x_i$ are
the universal quantities which depends on the BCs: for periodic
BCs ($\Delta = 0, x_1=1/8, x_2=1$), for antiperiodic BCs ($\Delta
= 1/16, x_1=3/8, x_2=1$), and for free BCs ($\Delta = 0, x_1=1/2,
x_2=2$).

Quite recently, Izmailian and Hu \cite{01prl} studied the finite
size correction terms for the free energy per spin and the inverse
correlation lengths of critical 2D Ising models  on $N \times
\infty$ lattice and 1D quantum Ising chain with periodic BCs. They
obtained analytic expressions for the finite-size correction
coefficients $a_k$, $b_k$ and $c_k$ in the expansions
\begin{eqnarray}
N [f(N)-f_{\infty}] &=&\sum_{k=1}^{\infty}\frac{a_k}{N^{2 k-1}},
\label{fN} \\
 \xi_s^{-1}(N)&=&\sum_{k=1}^{\infty}\frac{b_k}{N^{2
k-1}}, \label{cli} \\
\xi_e^{-1}(N) &=& \sum_{k=1}^{\infty}\frac{c_k}{N^{2 k-1}},
\label{clienergy}
\end{eqnarray}
and find that although the finite-size correction coefficients
$a_k$, $b_k$ and $c_k$ are not universal, the amplitude ratios for
the coefficients of these series are universal and given by
\begin{eqnarray}
r_s(k) &=& \frac{b_k}{a_k} = \frac{2^{2k}-1}{2^{2k-1}-1}, \label{I4}\\
r_e(k)&=&\frac{c_k}{a_k}=\frac{4 k}{(2^{2k-1}-1)B_{2k}},
\label{I5}
\end{eqnarray}
where $B_n$ is the $n$-th Bernoulli number ($B_2 = 1/6,B_4 =
-1/30, \dots$).

In the next section of the present paper we will present exact
calculations for a set of universal amplitude ratios for the 1D
quantum spin model \cite{katsura}, which is the quantum version of
the classical 2D Ising model, with free and antiperiodic BCs. We
obtain analytic equations for $a_k$, $b_k$ and $c_k$ in the
expansions given by Eqs. (\ref{fN}), (\ref{cli}) and
(\ref{clienergy}) and find that universal amplitude ratios for the
1D quantum Ising model with antiperiodic boundary condition are
given by
\begin{eqnarray}
r_s(k) &=& \frac{b_k}{a_k} = \frac{(2^{2k}-1)B_{2k}-2 k}{2^{2k-1}
B_{2k}}, \label{I8}\\
r_e(k) &=& \frac{c_k}{a_k} = -\frac{2 k}{B_{2k}}, \label{I9}
\end{eqnarray}
while for free BCs we obtain
\begin{eqnarray}
r_s(k) &=& \frac{b_k}{a_k} = \frac{4 k}{(2^{2k-1}-1)B_{2k}}, \label{I6}\\
r_e(k) &=& \frac{c_k}{a_k} = \frac{4 k
(3^{2k-1}+1)}{(2^{2k-1}-1)B_{2k}}. \label{I7}
\end{eqnarray}
 As far as we know, no previous
RG arguments, analytic calculations, or numerical studies predict
the existence of this whole set of universal amplitude ratios.

\section{One-dimensional quantum spin chain}

The 1D quantum spin chain, the hamiltonian limit of the classical
2D Ising model (see \cite{kogut} and references therein), belongs
to the most frequently investigated and best understood systems in
statistical physics. We will calculate universal amplitude ratios
for this systems with anti-periodic boundary conditions and free
boundary conditions.

\subsection{Anti-periodic boundary conditions}

Let us first consider the antiperiodic BCs.
On a chain with $N$ sites the Hamiltonian is given by
\begin{equation}
H=-\frac{\lambda}{2 \gamma}\sum_{n=1}^N \sigma^z_n -\frac{1}{4
\gamma}\sum_{n=1}^N \left[(1+\gamma)\sigma^x_{n+1}\sigma^x_n+
(1-\gamma)\sigma^y_{n+1}\sigma^y_n\right], \label{chain}
\end{equation}
which was exactly solved by Katsura \cite{katsura}. Here
$\sigma^x, \sigma^y$ and $\sigma^z$ are the Pauli spin matrices.
The antiperiodic BCs imposed on Eq. (\ref{chain}) are
$$
\sigma^x_{N+1}=-\sigma^x_{1} \qquad \mbox{and} \qquad
\sigma^y_{N+1}=-\sigma^y_{1}.
$$
The phase diagram is well known \cite{bm71}. For all $\gamma$
$(0<\gamma \le 1)$, there is a critical point at $\lambda_c =1$,
which falls into the 2D Ising universality class. For $\gamma = 1$
it is also called 1D transverse Ising model. Thus, by the
introduction of a parameter $\gamma$ we could study different
models in the same universality class. The Hamiltonian of Eq.
(\ref{chain}) can be diagonalized by a Jordan-Wigner
transformation as
\begin{equation}
H=\sum_k\Lambda_k\left(\eta_k^{\dag}\eta_k-1/2\right),\label{chain1}
\end{equation}
where $\eta_k^{\dag}, \eta_k$ are fermionic creation and
annihilation operators and
\begin{equation}
\Lambda_k=\sqrt{(\cos k-\lambda)^2/\gamma^2 + \sin^2
k},\label{Lambdak}
\end{equation}
is the lattice dispersion relation. At the critical point
$\lambda_c=1$ one then obtains $\Lambda_k=2
\psi\left(\frac{k\pi}{2N}\right)$. Here
\begin{equation}
\psi(x)=\sqrt{\sin^2(x)-\frac{\gamma^2-1}{\gamma^2} \sin^4
(x)}.\label{psi}
\end{equation}
The critical ground-state energy, $E_0^{(A)}$, corresponds to the
antiperiodic BC  has the value \cite{henkel87}
\begin{eqnarray}
E_0^{(A)} &=&-\sum_{m = 0}^{N-1}\psi\left(\frac{\pi m}{N}\right) =
-\sum_{m = 0}^{N-1}\sqrt{\sin^2 \frac{\pi
m}{N}-\frac{\gamma^2-1}{\gamma^2} \sin^4 \frac{\pi m}{N}}.
\label{groundantiperiodic}
\end{eqnarray}
The energy gaps $\Delta_s^{(A)}$ and $\Delta_e^{(A)}$ are given by
\begin{eqnarray}
\Delta_s^{(A)}&=& 2 \psi\left(\frac{\pi}{2 N}\right)+ \sum_{m =
0}^{N-1}\left[\psi\left(\frac{\pi m}{N}\right)-\psi\left(\frac{2
m +1}{2 N}\pi\right)\right], \label{antispinperiodic}\\
\Delta_e^{(A)}&=&2\psi\left(\frac{\pi}{N}\right)=
2\sqrt{\sin^2\frac{\pi}{N}-\frac{\gamma^2-1}{\gamma^2} \sin^4
\frac{\pi}{N}}. \label{energyspinperiodic}
\end{eqnarray}
Note, that the ground state energy $E_0$, the  first energy gap
($E_1-E_0 \equiv \Delta_s $) and the second energy gap ($E_2-E_0
\equiv \Delta_e$) of a quantum spin chain are, respectively, the
quantum analogies of the free energy $f(N)$, inverse spin-spin
correlation length $\xi_s^{-1}(N)$, and inverse energy-energy
correlation length $\xi_e^{-1}(N)$ for the Ising model; that is,
\begin{eqnarray}
N f(N) \Leftrightarrow - E_0, \; \xi_s^{-1}(N) \Leftrightarrow
\Delta_s, \; \mbox{and} \; \xi_e^{-1}(N) \Leftrightarrow \Delta_e.
\nonumber
\end{eqnarray}
To write $E_0^{(A)}$, $\Delta_s^{(A)}$, and $\Delta_e^{(A)}$ in
the form of Eqs. (\ref{fN}), (\ref{cli}), and (\ref{clienergy}),
we must evaluate Eqs. (\ref{groundantiperiodic}),
(\ref{antispinperiodic}), and (\ref{energyspinperiodic})
asymptotically. These sums can be handled by using the
Euler-Maclaurin summation formula \cite{hardy}. Suppose that
$F(x)$ together with its derivatives is continuous within the
interval $(a, b)$. Then the general Euler-Maclaurin summation
formula states
\begin{equation}
\sum_{n=0}^{N-1} F(a+n h+\alpha h)=\frac{1}{h}\int_{a}^{b}
F(\tau)~{\rm d}\tau + \sum_{k=1}^{\infty} \frac{h^{k-1}}{k!} {\rm
B}_{k}(\alpha)\left(F^{(k-1)}(b)-F^{(k-1)}(a)\right)
\label{EMFormula}
\end{equation}
where $0 \le \alpha \le 1$, $h=(b-a)/N$ and ${\rm B}_{k}(\alpha)$
are so-called Bernoulli polynomials defined in terms of the
Bernoulli numbers $B_p$ by
\begin{equation}
{\rm B}_{k}(\alpha)=\sum_{p=0}^{k}{\rm B}_{p}\frac{k!}{(k-p)!p!}
\alpha^{k-p}
\end{equation}
Indeed, $B_n(0)=B_n$. Bernoulli polynomials satisfy the identity:
\begin{equation}
B_n(1/2)=\left(2^{1-n}-1\right)B_n \label{bernoulli}
\end{equation}
By expanding the exact solution of Eq. (\ref{chain}), Henkel \cite{henkel87}
has obtained several finite-size correction terms to the ground-state energy
$E_0$ and to the first energy gap $E_1-E_0$. We have extended the
calculations to arbitrary order and found that
\begin{eqnarray}
E_0^{(A)}&+& N \alpha_0 = \sum_{k=1}^{\infty}\frac{2 B_{2 k}}{(2
k)!} \left(\frac{\pi}{N}\right)^{2 k-1}\psi^{(2 k-1)}
\nonumber \\
&=&\frac{\pi}{6 N}
-\frac{1}{15}\left(\frac{1}{\gamma^2}-\frac{4}{3}\right)
\left(\frac{\pi}{2 N}\right)^3+\dots, \label{E0}\\
\Delta_s^{(A)}&=& \sum_{k=1}^{\infty}\frac{4 k-2 B_{2 k}(2^{2
k}-1)}{(2 k)!} \left(\frac{\pi}{2 N}\right)^{2 k-1} \psi^{(2 k-1)}
\nonumber\\
&=&\frac{3\pi}{4
N}+\frac{9}{8}\left(\frac{1}{\gamma^2}-\frac{4}{3}\right)
\left(\frac{\pi}{2 N}\right)^3+\dots,
\label{E10}\\
\Delta_e^{(A)}&=& \sum_{k=1}^{\infty}\frac{4 k}{(2 k)!}
\left(\frac{\pi}{N}\right)^{2 k-1} \psi^{(2 k-1)}
\nonumber\\
&=&\frac{2 \pi}{N}+\left(\frac{1}{\gamma^2}-\frac{4}{3}\right)
\left(\frac{\pi}{N}\right)^3+\dots, \label{enerener2}
\end{eqnarray}
where  $\psi^{(2 k-1)} = \left(d^{2 k-1} \psi(x)/dx^{2
k-1}\right)_{x=0}$ and $\alpha_0$ is an non-universal number
\begin{equation}\alpha_0 = \frac{1}{\pi}\int_0^{\pi} \psi(x) d x =
\left[1+\arccos{\gamma}/(\gamma \sqrt{1-\gamma^2})\right]/\pi.
\label{alpha0}
\end{equation}
The ratios of the amplitudes of the $N^{-(2k-1)}$ correction terms
in the spin-spin correlation length, energy-energy correlation
length and the free energy expansion, {\it i.e.} $b_k/a_k$ and
$c_k/a_k$, are $\gamma$-independent and are given by Eqs.
(\ref{I8}) and (\ref{I9}), respectively, thus confirming the
universality of this ratios.

\subsection{Free boundary conditions}

Let us now consider the transverse Ising model on a 1D lattice of
$N$ sites with free BC and with the Hamiltonian
\begin{equation}
H = -\lambda \sum_{n=1}^N \sigma_n^z - 2 \sum_{n=1}^{N-1}
\sigma_n^x \sigma_{n+1}^x. \label{hamiltonian}
\end{equation}
In the limit $N \to \infty$, the ground state is singular at
$\lambda = 1$. The Ising chain in a transverse field has been
studied in great detail (see e.g. \cite{katsura,pfeuty}). The
Hamiltonian of Eq. (\ref{hamiltonian}) may be re-expressed in the
diagonal form
\begin{equation}
H=\sum_k \bar
\Lambda_k\left(\eta_k^{\dag}\eta_k-1/2\right)\label{hamiltonian1}
\end{equation}
 with dispersion relation
\begin{equation}
\bar \Lambda_k= \sqrt{(\lambda-1)^2+4 \lambda
\sin^2(k/2)}.\label{barlambdak}
\end{equation}
For free BCs one finds that the allowed values of $k$ are
determined by the secular equation \cite{pfeuty}
\begin{equation}
\lambda^{-1}=\sin[(N+1)k]/\sin(N k). \label{lambda-1}
\end{equation}
At the critical field $\lambda_c = 1$, the secular
equation reduced to
\begin{equation}
\tan(k N)=\cot(k/2),\label{seculat}\end{equation}
 so that
\begin{equation}
k=\frac{(2 m+1)\pi}{2 N+1}, \hspace{1cm} m = 0, 1, \dots, N-1.
\nonumber
\end{equation}
and for the dispersion relation one then obtain $\bar \Lambda_k=2
\varphi \left(\frac{(k+1/2)\pi}{2N+1}\right)$, where
\begin{equation}
\varphi(x)=\sin(x).\label{varphi}
\end{equation}
The ground-state energy $E_0^{(F)}$, the energy gaps
$\Delta_s^{(F)}$ and $\Delta_e^{(F)}$ are given by \cite{pfeuty}
\begin{eqnarray}
E_0^{(F)} &=& - \sum_{m = 0}^{N-1}\varphi\left(\frac{m
+\frac{1}{2}}{2 N+1}\pi\right)=\frac{1}{2}\left(1-{\rm
cosec}\frac{\pi}{4 N+2}\right), \\
\Delta_s^{(F)}&=&2\varphi\left(\frac{\pi}{4N+2}\right)=2
\sin\frac{\pi}{2 (2 N+1)},
 \\
\Delta_e^{(F)}&=&2\left( \sin\frac{\pi}{2 (2 N+1)}+\sin\frac{3
\pi}{2 (2 N+1)}\right).
\end{eqnarray}
The asymptotic expansion of these quantities can be written in the
following form
\begin{eqnarray}
E_0^{(F)}&+& N \alpha_0 = \sum_{k=1}^{\infty}\frac{B_{2 k}(1-2^{2
k-1}}{(2 k)!} \left(\frac{\pi}{4 N+2}\right)^{2 k-1}\varphi^{(2
k-1)} \nonumber\\
&=&-\frac{\pi}{24(2 N+1)} -\frac{7}{90} \left(\frac{\pi}{8
N+4}\right)^3+\dots, \label{fN-all1}\\
\Delta_s^{(F)} &=& \sum_{k=1}^{\infty}\frac{2}{(2 k - 1)!}
\left(\frac{\pi}{4 N+2}\right)^{2 k-1} \varphi^{(2 k-1)}
\nonumber\\
&=&\frac{\pi}{2N+1} -\frac{1}{3} \left(\frac{\pi}{4
N+2}\right)^3+\dots,
\label{I191}\\
\Delta_e^{(F)} &=& \sum_{k=1}^{\infty}\frac{2(3^{2k-1}+1)}{(2
k-1)!} \left(\frac{\pi}{4 N+2}\right)^{2 k-1} \varphi^{(2 k-1)}
\nonumber\\
&=&\frac{4\pi}{2 N+1} -\frac{28}{3} \left(\frac{\pi}{4
N+2}\right)^3+\dots, \label{I181}
\end{eqnarray}
where,  $\varphi^{(2 k-1)} = \left(d^{2 k-1} \varphi(x)/dx^{2
k-1}\right)_{x=0}$, $\varphi(x)= \sin x$, and  $\alpha_0 =
\frac{1}{\pi}\int_0^{\pi} \varphi(x) d x = 2/\pi$.

Equations (\ref{fN-all1}), (\ref{I191}), and (\ref{I181}) imply
that the ratios of the amplitudes of the $N^{-(2k-1)}$ correction
terms in the spin-spin correlation length, the energy-energy
correlation lengths, and the free energy expansion, {\it i.e.}
$b_k/a_k$ and $c_k/a_k$, should not depend in detail on the
dispersion relation as given by Eqs. (\ref{I6}) and (\ref{I7}).

The leading terms of Eqs. (\ref{E0}) - (\ref{enerener2}),
(\ref{fN-all1}) - (\ref{I181})  are consistent with Eqs.
(\ref{I2}) - (\ref{Afree}), {\it i.e.} $a_1$, $b_1$ and $c_1$ are
universal. Equations (\ref{I2}) and (\ref{I1}) implies immediately
that their ratio is also universal, namely $r_s(1)=D_1/A$ and
$r_e(1)=D_2/A$, which is consistent with Eqs. (\ref{I4}),
(\ref{I8}), and Eq. (\ref{I6}) for the case $k=1$
\begin{eqnarray}
r_s(1)=\frac{D_1}{A} =\left\{\begin{array}{lcl}
3 \qquad \qquad \quad \mbox{for periodic BC} \label{rsperiod1}\\
-9/2 \quad \qquad \mbox{for antiperiodic BC} \label{rsantiperiod1}\\
24 \qquad \qquad \mbox{for free BC} \label{rsfree1}
\end{array}
\right.
\end{eqnarray}
and with Eqs. (\ref{I5}), (\ref{I9}), and Eq. (\ref{I7}) for the
case $k=1$
\begin{eqnarray}
r_e(1)=\frac{D_2}{A}=\left\{\begin{array}{lcl}
24 \qquad \qquad \quad \mbox{for periodic BC} \label{reperiod1}\\
-12\qquad \qquad \mbox{for antiperiodic BC} \label{reantiperiod1}\\
96 \qquad \qquad \quad\mbox{for free BC} \label{refree1}
\end{array}
\right.
\end{eqnarray}

\section{Perturbative  conformal field theory}

The finite-size corrections to Eqs. (\ref{I2}) and (\ref{I1}) can
be calculated by the means of a perturbative  conformal field
theory \cite{cardy86,zamol87}. In general, any lattice Hamiltonian
will contain correction terms to the critical Hamiltonian $H_c$
\begin{equation}
H = H_c + \sum_p g_p \int_{-N/2}^{N/2}\phi_p(v) d v, \label{Hc}
\end{equation}
where $g_p$ is a non-universal constant and $\phi_p(v)$ is a
perturbative  conformal field. Below we will consider the case
with only one perturbative  conformal field, say $\phi_l(v)$. Then
the eigenvalues of $H$ are
\begin{equation}
E_n=E_{n,c}+ g_l \int_{-N/2}^{N/2}<n|\phi_l(v)|n> d v +
\dots,\label{En}
\end{equation}
where $E_{n,c}$ are the critical
eigenvalues of $H$. The matrix element $<n|\phi_l(v)|n>$ can be
computed in terms of the universal structure constants $(C_{nln})$
of the operator product expansion \cite{cardy86}: $<n|\phi_l(v)|n>
=\left({2 \pi}/{N}\right)^{x_l}C_{nln}$, where $x_l$ is the
scaling dimension of the conformal field $\phi_l(v)$. The energy
gaps $(\Delta_n= E_n-E_0)$ and the ground-state energy ($E_0$) can
be written as
\begin{eqnarray}
\Delta_n&=&\frac{2 \pi}{N} x_n+ 2 \pi
g_l(C_{nln}-C_{0l0})\left(\frac{2 \pi}{N}\right)^{x_l-1} + \dots,
\label{xin}\\
E_0 &=& E_{0,c}+2 \pi  g_l C_{0l0} \left(\frac{2
\pi}{N}\right)^{x_l-1} + \dots. \label{E0conf}
\end{eqnarray}
For the 2D Ising model, one finds \cite{01prl} that at least two
(and probably infinitely many) perturbative  conformal fields are
necessary to generate all finite-size corrections terms.
Nevertheless, the leading finite-size corrections ($1/N^3$) can be
described by the Hamiltonian given by Eq. (\ref{Hc}) with a single
perturbative conformal field $\phi_l(v)=L_{-2}^2(v)+{\bar
L}_{-2}^2(v)$ with scaling dimension $x_l=4$ \cite{henkel}.

In order to obtain the corrections we need the matrix elements
$<n|L_{-2}^2(v)+{\bar L}_{-2}^2(v)|n>$, which have already been
computed by Reinicke \cite{reinicke87}:
\begin{eqnarray}
<\Delta+r|L_{-2}^2|\Delta+r> &=&
\left(\frac{2\pi}{N}\right)^4\left[\frac{49}{11520}+(\Delta+r)
\left(\Delta-\frac{5}{24}+\frac{r(2
\Delta + r)(5 \Delta+1)}{(\Delta+1)(2\Delta+1)}\right)\right],
\label{L-2Delta}\\
<r|L_{-2}^2|r> &=&
\left(\frac{2\pi}{N}\right)^4\left[\frac{49}{11520}+\frac{49}{120}r(2r^2-3)\right].
\label{L-2}
\end{eqnarray}
The universal structure constants $C_{2l2}$, $C_{1l1}$ and
$C_{0l0}$ can be obtained from the matrix element
\begin{equation}
<n|L_{-2}^2(v)+{\bar L}_{-2}^2(v)|n> =\left({2
\pi}/{N}\right)^{x_l}C_{nln},\label{L22}
\end{equation}
where $x_l = 4$ is the scaling dimension of the conformal field
$L_{-2}^2(v)+{\bar L}_{-2}^2(v)$.

At the critical point $\lambda_c=1$ the spectra of the Hamiltonian
(\ref{chain}) with periodic and antiperiodic BC are built by the
irreducible representation $\Delta, \bar \Delta$ of two commuting
Virasoro algebras $L_n$ and ${\bar L}_n$ with central charge
$c=\frac{1}{2}$ \cite{cardy86a}. We denote by $\Delta$ the highest
weight, and by $\Delta+r$, the $r$-th level having degeneracy
$d(\Delta,r)$ of one irreducible representation of the Virasoro
algebra. A state will be labelled by $|n> \sim |\Delta+r, \bar
\Delta+\bar r>$. The possible values of $\Delta, \bar \Delta$ are
$(0,0), (\frac{1}{2},\frac{1}{2}),(\frac{1}{16},\frac{1}{16})$ for
periodic and
$(0,\frac{1}{2}),(\frac{1}{2},0),(\frac{1}{16},\frac{1}{16})$ for
antiperiodic BCs. In the case of free BCs the spectra can be
understood in terms of irreducible representations $\Delta$ of a
single Virasoro algebra with possible values of $\Delta$ are $0,
 \frac{1}{2}$.

The ground state $|0>$, first excited state $|1>$, and second
excited state $|2>$ depends on the boundary conditions and given
by \cite{cardy86,gehlen}:
\begin{eqnarray}|0> &=& |\Delta=0,r=0;\bar
\Delta=0,\bar r=0>, \label{state0p}\\ |1> &=&
|\Delta=\frac{1}{16},r=0;\bar \Delta=\frac{1}{16},\bar r=0>,
\label{state1p}\\ |2> &=& |\Delta=\frac{1}{2},r=0;\bar
\Delta=\frac{1}{2},r=0>,\label{state2p}
\end{eqnarray}
for periodic BCs;
\begin{eqnarray}
|0>&=&|\Delta=\frac{1}{16},r=0;\bar \Delta=\frac{1}{16},\bar
r=0>,\label{stae0ap}\\ |1>&=&|\Delta=0,r=0;\bar
\Delta=\frac{1}{2},\bar r=0>, \label{state1ap}\\
|2>&=&|\Delta=\frac{1}{16},r=0;\bar \Delta=\frac{1}{16},\bar
r=1>,\label{state2ap}\end{eqnarray} for antiperiodic BCs; and
\begin{eqnarray}
|0>&=&|\Delta=0,r=0>, \label{state0f}\\
|1>&=&|\Delta=\frac{1}{2},r=0>, \label{state1f}\\
|2>&=&|\Delta=0,r=2>\label{state2f}
\end{eqnarray}
for free BCs.

After reaching this point, one can easily compute the universal
structure constants $C_{0l0}$, $C_{1l1}$ and $C_{2l2}$ for all
three boundary conditions. The values of $C_{0l0}$, $C_{1l1}$,
$C_{2l2}$ can be obtained from Eqs. (\ref{L-2Delta}) -
(\ref{state2f}) and given by:
\begin{equation}
C_{0l0}=49/5760, \qquad C_{1l1} = -7/720, \qquad
C_{2l2}=1729/5760,\label{Cp}
\end{equation}
for periodic BCs;
\begin{equation}
C_{0l0}=-7/720, \qquad C_{1l1} = 889/5760, \qquad
C_{2l2}=833/720,\label{Ca}
\end{equation}
for antiperiodic BCs; and
\begin{equation}
C_{0l0}=49/11520, \qquad C_{1l1} = 1729/11520, \qquad
C_{2l2}=47089/11520,\label{Cf}
\end{equation}
for free BCs. Equations (\ref{xin}) and (\ref{E0conf}) implies
that the ratios of first-order corrections amplitudes for
($\Delta_n$) and ($- E_0$) is universal and equal to
$(C_{0l0}-C_{nln})/C_{0l0}$, which is consistent with Eqs.
(\ref{I4}), (\ref{I8}), and Eq. (\ref{I6}) for the case $k=2$
\begin{eqnarray}
r_s(2)=\frac{C_{0l0}-C_{1l1}}{C_{0l0}} =\left\{\begin{array}{lcl}
15/7 \qquad \quad \mbox{for periodic BCs,} \label{rsperiod}\\
135/8 \quad \quad \mbox{for antiperiodic BCs,} \label{rsantiperiod}\\
-240/7 \qquad \mbox{for free BCs} \label{rsfree}
\end{array}
\right.
\end{eqnarray}
and with Eqs. (\ref{I5}), (\ref{I9}), and Eq. (\ref{I7}) for the
case $k=2$
\begin{eqnarray}
r_e(2)=\frac{C_{0l0}-C_{2l2}}{C_{0l0}}=\left\{\begin{array}{lcl}
-240/7 \quad \mbox{for periodic BCs,} \label{reperiod}\\
120 \qquad \quad \mbox{for antiperiodic BCs,} \label{reantiperiod}\\
-960 \qquad \mbox{for free BCs.} \label{refree}
\end{array}
\right.
\end{eqnarray}

\section{Conclusion}

In this paper we discuss the influence of the boundary conditions
on the finite-size corrections of the 2D Ising model in the
extreme anisotropic or quantum-Hamiltonian limit. We have
calculated various universal amplitude ratios and find that such
result are in perfect agreement with a perturbated conformal field
theory scenario proposed by Cardy \cite{cardy86}.

The results of this paper inspire several problems for further
studies: (i) Further work has to be done to possibly evaluate
exactly all finite-size correction terms from perturbative
conformal field theory. (ii) Can one obtain from the perturbated
conformal field theory the value of the universal amplitude ratios
($r_s(k)$ and $r_e(k)$) for $k > 2$? (iii) How do such amplitudes
behave in other models, for example in the three-state Potts
model? (iv) Our results also present new challenges to scientists
working on numerical studies of critical phenomena. For example,
it is of interest to present accurate numerical evidences about
whether the Ising model on a two-dimensional lattice with crossing
bonds has the same set of amplitude ratios.

\section{Acknowledgements}
One of us (NSI) would like to thank A. Zamolodchikov for helpful
comments and discussions. This work was supported by National
Science Council of the Republic of China (Taiwan) under Grant No.
NSC 96-2911-M 001-003-MY3 and National Center for Theoretical
Sciences in Taiwan.

%%%%%%%%%%%%%%%%%%%%%%%%%%%%%%%%%%%%%%%%%%%%%%%%%%%%%%%%%%%%%%%%%%%%%%%%%%

%%%%%%%%%%%%%%%%%%%%%%%%%%%%%%%%%%%%%%%%%%%%%%%%%%%%%%%%%%%%%%%%%%%%%%%%%%%

\end{document}